\documentclass[5p,preprint]{elsarticle}
 
\usepackage[utf8]{inputenc}
\usepackage{siunitx}
\usepackage{graphicx}
\usepackage{amsmath,mathtools, cuted}
\usepackage{subcaption}
\usepackage[table,dvipsnames]{xcolor}
\usepackage{booktabs}
\usepackage{multirow}
\usepackage{bm}
\usepackage{textcomp}
\usepackage{float}
\usepackage{gensymb}
\usepackage{cleveref}
\usepackage{chemformula}
\usepackage{amsopn}
\usepackage{tabularx}

\journal{Acta Materialia}

\begin{document}

\begin{frontmatter}

\title{Accelerating CALPHAD-based Phase Diagram Predictions in Complex Alloys Using Universal Machine Learning Potentials: Opportunities and Challenges}

\author[1]{Siya Zhu}
\ead{siyazhu@tamu.edu}

\author[1]{Doguhan Sariturk}
\ead{sariturk@tamu.edu}

\author[1,2,3]{Raymundo Arr\'oyave}
\ead{raymundo.arroyave@tamu.edu}

\affiliation[1]{organization={Department of Materials Science and Engineering, Texas A\&M University},
                city={College Station},
                postcode={77843}, 
                state={TX},
                country={USA}}

\affiliation[2]{organization={J. Mike Walker '66 Department of Mechanical Engineering, Texas A\&M University},
                city={College Station},
                postcode={77843}, 
                state={TX},
                country={USA}}

\affiliation[3]{organization={Wm Michael Barnes '64 Department of Industrial and Systems Engineering, Texas A\&M University},
                city={College Station},
                postcode={77843}, 
                state={TX},
                country={USA}}

\begin{abstract}
Accurate phase diagram prediction is crucial for understanding alloy thermodynamics and advancing materials design. While traditional CALPHAD methods are robust, they are resource-intensive and limited by experimentally assessed data. This work explores the use of machine learning interatomic potentials (MLIPs) such as M3GNet, CHGNet, MACE, SevenNet, and ORB to significantly accelerate phase diagram calculations by using the \emph{Alloy Theoretic Automated Toolkit} (ATAT) to map calculations of the energies and free energies of atomistic systems to CALPHAD-compatible thermodynamic descriptions. Using case studies including \ch{Cr-Mo}, \ch{Cu-Au}, and \ch{Pt-W}, we demonstrate that MLIPs, particularly ORB, achieve computational speedups exceeding three orders of magnitude compared to DFT while maintaining phase stability predictions within acceptable accuracy. Extending this approach to liquid phases and ternary systems like \ch{Cr-Mo-V} highlights its versatility for high-entropy alloys and complex chemical spaces. This work demonstrates that MLIPs, integrated with tools like ATAT within a CALPHAD framework, provide an efficient and accurate framework for high-throughput thermodynamic modeling, enabling rapid exploration of novel alloy systems. While many challenges remain to be addressed, the accuracy of some of these MLIPs (ORB in particular) are on the verge of paving the way toward high-throughput generation of CALPHAD thermodynamic descriptions of multi-component, multi-phase alloy systems.
\end{abstract}


\end{frontmatter}

\section{Introduction}
Phase diagram calculations are of utmost importance in studying the thermodynamic properties of alloys, which in turn constitute the foundation for any deep understanding of materials behavior. The CALPHAD (CALculation of PHAse Diagrams) method \cite{lukas2007computational}---implemented in software packages such as OpenCalphad \cite{sundman2015opencalphad}, Thermo-Calc \cite{sundman1985thermo,andersson2002thermo}, and Pandat \cite{chen2002pandat}---is a computational approach used to model and predict phase equilibria and thermodynamic properties in multicomponent systems. Developed in the 1970s \cite{kaufman1973theoretical} the CALPHAD framework involves constructing mathematical models to describe the Gibbs free energy of each phase as a function of temperature, pressure, and composition, and investigate the thermodynamic equilibria among different phases \cite{saunders1998calphad,liu2009first,lukas2007computational}. Since its inception, the CALPHAD method was designed as a hierarchical system whereby low-order (e.g. unary, binary) thermodynamic descriptions could be expanded to encompass larger numbers of constituents and phases, enabling the development of increasingly comprehensive thermodynamic databases. Starting from reference thermodynamic descriptions of the pure elements standardized and disseminated through the SGTE (Scientific Group Thermodata Europe) elemental database \cite{dinsdale1991sgte} the CALPHAD community has been able to construct highly complex databases that encompass dozens of constituents and hundreds of phases---such as Thermo-Calc's High Entropy Alloy database \cite{mao2017tchea1}---and that are now widely used to predict the phase stability and thermo-physical properties of materials, using in turn these calculations in complex materials design workflows \cite{arroyave2022perspective,arroyave2019systems}. 

Despite the promise of CALPHAD-based methods, there remain important limitations. CALPHAD-based assessment of thermodynamic databases is exceedingly resource-intensive and most of the fully assessed systems correspond to binary alloys, with a modest (albeit growing) number of ternary thermodynamic descriptions being added to CALPHAD databases every year. A case in point correspond to the high entropy alloy (HEA) space \cite{zhu2023probing,nataraj2021systematic,nataraj2021temperature}, as this new paradigm in alloy design has encouraged the community to probe regions in the alloy chemistry space that have been seldom investigated from a phase stability perspective. As the materials community sets its sights on compositionally complex chemical spaces (beyond HEAs), a major challenge is the fact that CALPHAD methods are limited in their predictive ability as they can only predict the phase stability of phases that have been observed experimentally and at least partially assessed. This is an important limitation when attempting to investigate the phase stability of systems that are not available in commercial or even open databases \cite{van2018thermodynamic}.

To overcome the limitations of traditional approaches, computational tools have been developed to streamline the process of generating CALPHAD databases from \textit{ab initio} data. Among these, the Alloy Theoretic Automated Toolkit (ATAT) \cite{van2002alloy,van2017software} stands out as a versatile framework for bridging the gap between \textit{ab initio} computational data and thermodynamic models. ATAT is particularly well-suited for workflows involving density functional theory (DFT) calculations performed using widely adopted software such as VASP \cite{hafner2008ab}. The toolkit provides systematic methods for converting formation energies and other thermodynamic properties derived from \textit{ab initio} calculations into formats compatible with CALPHAD-based thermodynamic modeling. 

One of ATAT's core features is the leveraging of the Special Quasirandom Structures (SQS) \cite{zunger1990special} framework. This methodology facilitates the (approximate) simulation of disordered atomic arrangements within various crystal systems. While traditional applications often focus on common structures such as FCC, BCC, and HCP, ATAT extends this capability to handle complex multiple-sublattice systems, including those with sublattice disorder. To enhance usability, ATAT includes a comprehensive pre-generated database of SQS configurations. A critical advantage of ATAT is its ability to address the challenge of assigning free energies to mechanically unstable "virtual" phases that are ill-defined from a rigorous thermodynamic standpoint. By using a theoretically robust approach, ATAT enables the inclusion of these phases in CALPHAD models, addressing a common hurdle when extrapolating \textit{ab initio} data to practical thermodynamic applications. Additionally, ATAT incorporates a low-order approximation scheme to account for short-range order effects, eliminating the need for additional computational inputs while maintaining model accuracy. The integration of ATAT with the Scientific Group Thermodata Europe (SGTE) elemental databases further enhances its utility. This allows ATAT to seamlessly merge \textit{ab initio} formation energies and vibrational free energies with established thermodynamic datasets, resulting in complete and functional thermodynamic models. These models can be directly employed in CALPHAD workflows, facilitating high-throughput exploration and rigorous thermodynamic assessments.

While the use of \textit{ab initio} calculated energetics of Special Quasirandom Structures (SQS) for developing CALPHAD models is effective for exploring large chemical-phase spaces, other tools are better suited for studying the detailed thermodynamic properties of single phases with significant solid solubility. One such tool is the Cluster Expansion (CE) method \cite{sanchez1984generalized,de1994cluster,zunger1994first}.

The CE method approximates the total energy of substitutional alloys by representing interactions among clusters of atoms. These clusters include single atoms, pairs, triplets, and higher-order groupings, all within a consistent parent lattice defined by its symmetry. This framework makes it possible to explore compositional variations systematically and investigate properties such as local ordering and phase segregation. The CE method is often combined with semi-grand canonical Monte Carlo simulations for further analysis \cite{connolly1983density,van2002automating,van2009multicomponent}. A major advantage of the CE method is its ability to represent complex alloy configurations more efficiently than direct Density Functional Theory (DFT) calculations for large systems. The interaction terms in CE are derived by fitting to DFT-calculated energies \cite{kresse1993ab,kresse1994ab,kresse1996efficiency,kresse1996efficient}, allowing accurate predictions of the energy for structures with hundreds of atoms. This makes CE particularly useful for studying disordered materials and substitutional alloys.

However, the CE method has limitations. It requires a large amount of training data to create models tailored to specific elements and phases, which can be time-consuming and computationally expensive. As the number of elements increases, the complexity grows significantly, because more clusters and interactions need to be accounted for \cite{kormann2017long}. For example, in the work by Nataraj \emph{et al.} \cite{nataraj2021systematic,nataraj2021temperature}, CE models for the NbTiVZr, HfNbTaTiZr, and AlTiZrNbHfTa alloy systems required fitting to 2,984, 1,970, and 4,000 structures, respectively. Each structure required DFT calculations, with each one taking 100 to 1,000 core-hours on a high-performance computing system. This computational demand makes CE models difficult to generalize or reuse across different systems. Any changes in composition require retraining the model, which is both labor-intensive and costly. While CE provides valuable insights into alloy behavior, its use for exploring multicomponent systems is limited by the trade-offs between required accuracy and available computational resources.

To accelerate the calculation of phase diagrams, machine learning interatomic potentials (MLIPs) have been utilized in several recent studies \cite{rosenbrock2021machine,hodapp2021machine}. Pre-trained universal MLIPs based on Graph Neural Network (GNN) architectures \cite{scarselli2008graph}, such as M3GNet \cite{chen2022universal}, CHGNet \cite{deng2023chgnet}, MACE \cite{Batatia2022mace,Batatia2022Design}, SevenNet \cite{park2024scalable,batzner20223}, and ORB \cite{neumann2024orb}, have demonstrated significant improvements in computational efficiency for structural optimization and free energy calculations compared to first-principles methods. These models reduce the computational costs associated with such calculations, albeit with a trade-off in accuracy, which is often moderate. Recent work by G. Vazquez \cite{vazquez2024deciphering} illustrates how machine learning can accelerate the Cluster Expansion (CE) process, particularly in the \ch{AlHfNbTaTiZr} high-entropy alloy (HEA) system. Similarly, MLIPs have been successfully applied as substitutes for DFT calculations in Special Quasirandom Structures (SQS)-based modeling \cite{hodapp2021machine,levamaki2022predicting}, demonstrating their potential in high-throughput thermodynamic analysis.

However, while MLIPs provide notable computational advantages, their application to phase diagram generation presents certain limitations. Discrepancies in equilibrium temperatures and phase topologies have been observed when comparing MLIP-derived thermodynamic data with experimental and \textit{ab initio} results. These discrepancies highlight the impact of small inaccuracies in MLIP-predicted energetics, which can propagate through calculations and lead to deviations in the predicted phase behavior. Such limitations underscore the importance of balancing efficiency with the level of precision required for reliable thermodynamic modeling.

\begin{figure}[hbt]
    \centering
        \includegraphics[width=\linewidth]{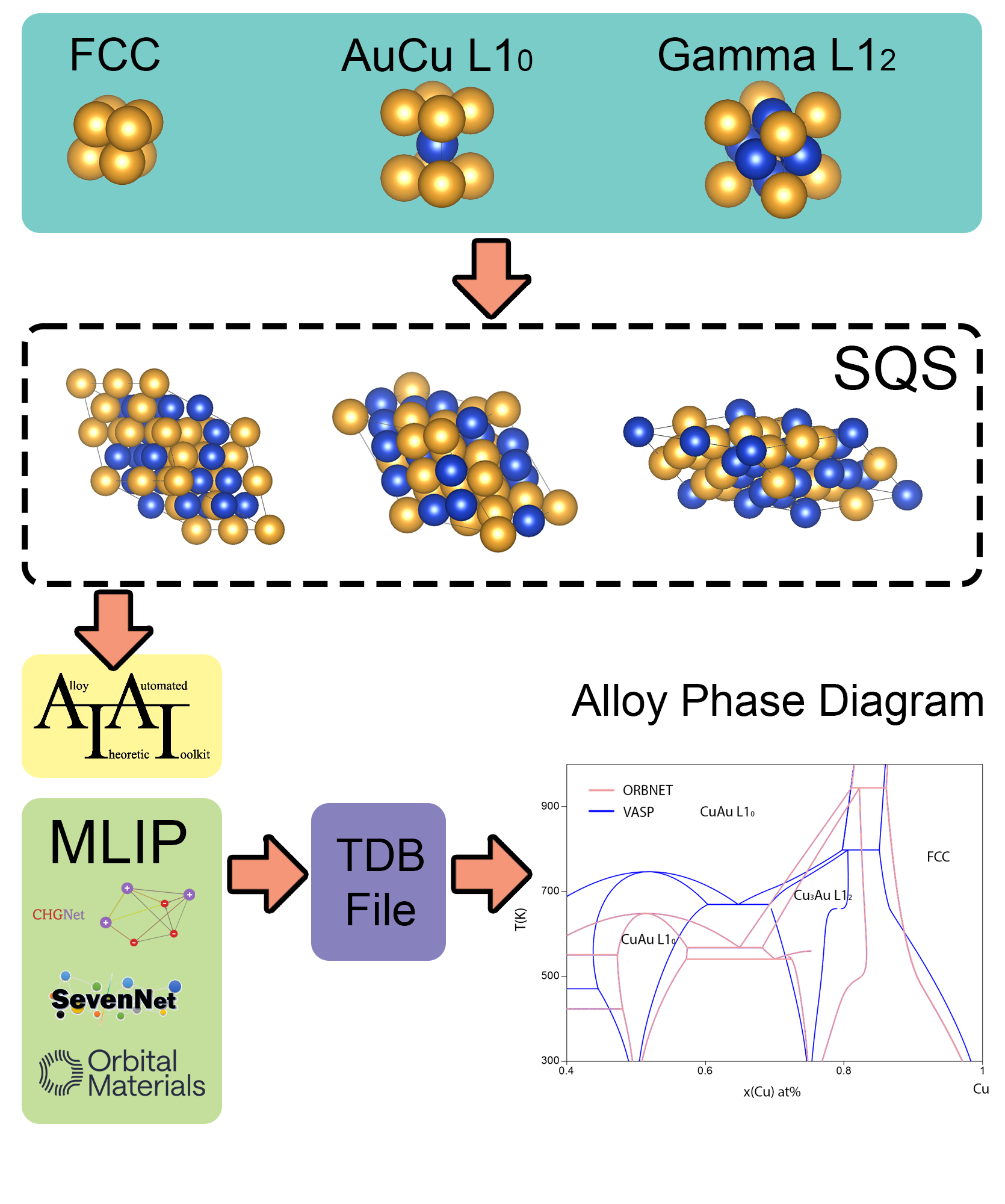}
    \caption{An example of how we calculate the Au-Cu phase diagram. Using the computational framework in ATAT, we generate SQSs for different phases and concentrations, and calculate the free energy using MLIPs. A TDB file is then generated, and the phase diagram is plotted using CALPHAD modeling.}
    \label{fig:abstract}
\end{figure}

In this work, using the computational framework employed in ATAT, we calculate alloy phase diagrams with MLIPs (with a workflow schematically shown in \Cref{fig:abstract}) and explore how the seemingly minor inaccuracies of MLIPs can cumulatively affect phase diagram calculations. We compare results obtained from M3GNet, CHGNet, MACE, SevenNet, and ORB across various alloy systems to quantify their accuracy. Based on these comparisons, we propose thresholds for the precision required in MLIPs to achieve reliable phase diagram predictions. Furthermore, we present example phase diagrams calculated using ORB, a universal MLIP that we find aligns closely with \textit{ab initio} results for several alloy systems. These examples demonstrate the potential of combining SQS and MLIPs for high-throughput alloy phase diagram calculations, offering a roadmap for integrating these approaches into future research and applications, including the possibility of generating CALPHAD databases at \textit{ab initio}-level accuracy---albeit still quantitatively apart from experimentally-derived CALPHAD descriptions---in a high-throughput manner, achieving computational efficiencies that are orders of magnitude greater than those possible with direct DFT calculations. 

\section{Methods}

For the CALPHAD modeling of formation free energy, we rely on the compound energy formalism \cite{hillert2001compound} applied in the standard TDB format. Following the work by van de Walle \emph{et al.} \cite{van2017software}, the Gibbs free energy of phase $\beta$ at site fraction $y$ and temperature $T$ is calculated as:

\begin{equation}
\begin{aligned}
  G^\beta\left(y,T\right)=\left(G^\beta_{\text{calc,nc}}\left(y,T\right) -\sum_i x_i G^{\alpha_i}_{\text{calc,nc}}\left(T\right)\right) \\
    + \sum_i x_i G^{\alpha_i}_{\text{SGTE}}\left(T\right) - T S_{\text{id}}\left(y,T\right)
\end{aligned}
\end{equation}

where the "calc,nc" stands for calculated Gibbs free energy  with no concentration entropy, and SGTE stands for the data from the SGTE database. SQSs are generated with the ATAT mcsqs code \cite{van2009multicomponent}, and the free energies of the SQSs at \SI{0}{\kelvin} are calculated with either \textit{ab initio} calculations or MLIPs. The interaction parameters are assumed temperature-independent and fitted within the regular solution model framework from the calculated energies of the generated SQS.

For the entropy calculations, configurational entropy is modeled using an ideal solution model on each sublattice, and the vibrational contributions are modeled using the Born von Karman model and were constructed with the fitfc code in ATAT package at the high-temperature limit \cite{van2002automating}, assuming that all phonon modes are excited beyond the Debye temperature.

For all the \textit{ab initio} calculations, the Vienna Ab Initio Simulation Package (VASP) \cite{kresse1993ab,kresse1994ab,kresse1996efficiency,kresse1996efficient} is used with the Perdew-Burke-Ernzerhof (PBE) exchange-correlation functional and PAW pseudopotentials at the level of the generalized gradient approximation (GGA) \cite{perdew1996generalized,blochl1994projector}. A K-points density of 8,000 k-points per reciprocal atom is set for all calculations. \par

In this study, pre-trained MLIPs were employed without additional fine-tuning to perform structural relaxations, single-point energy calculations, and molecular dynamics (MD) simulations. The models used included M3GNet (M3GNet-MP-2021.2.8-PES) \cite{chen2022universal}, CHGNet (0.3.0) \cite{deng2023chgnet}, MACE (MACE-MP-0 large) \cite{Batatia2022mace, Batatia2022Design}, SevenNet-0 (July 11, 2024) \cite{park2024scalable, batzner20223}, and ORB (orb-v2) \cite{neumann2024orb}. Structural relaxations were performed using the FIRE algorithm \cite{Bitzek2006FIRE} implemented in the Atomic Simulation Environment (ASE) \cite{ASE}.

MD simulations are performed for the average free energy for the liquid phase using the ASE framework \cite{ASE} with Nosé-Hoover ensembles \cite{Simone1993hoover}. We first generated liquid SQS configurations using ATAT at various concentrations and built a $2\times2\times2$ supercell. For the liquid phase MD simulations, we employ the NPT ensemble for the first 3,000 steps with a time step of \SI{1}{\femto\second} to obtain an appropriate structure and cell size. This is followed by 10,000 steps in the NVT ensemble to compute the average total energy. The temperature is set \SI{50}{\kelvin} above the melting points of the end members, with linear interpolation applied for the mixed structures.
\section{Results and Discussions}

\begin{figure}[hbt]
    \centering
    \includegraphics[width=\columnwidth]{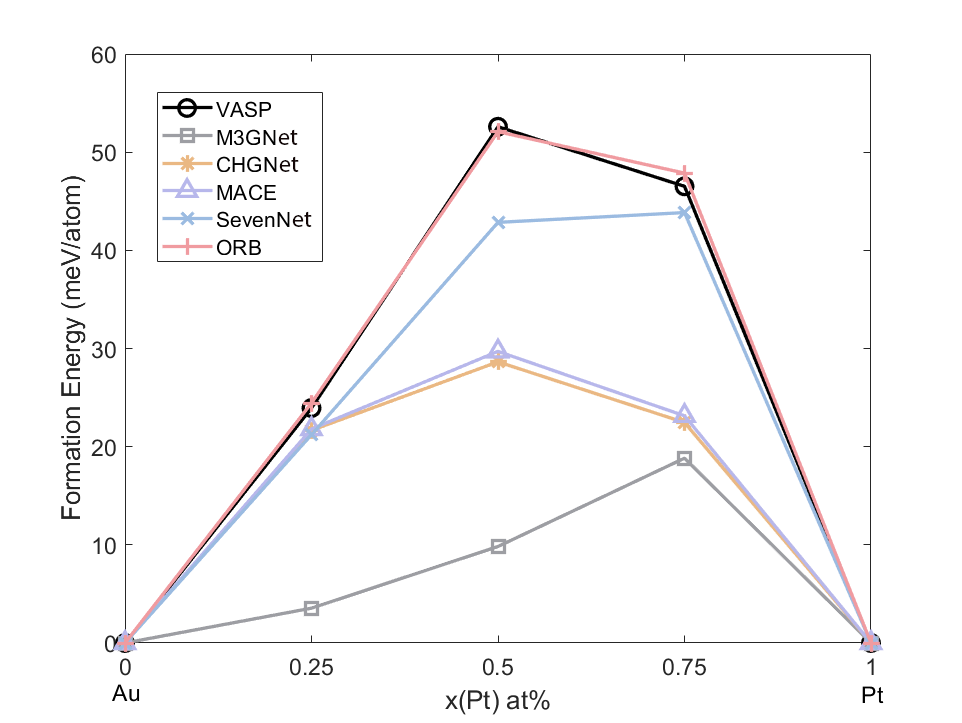}
	\caption{The formation energies of FCC \ch{Au-Pt} SQS calculated with VASP and MLIP data.}
	\label{fig:AuPtenergy}
\end{figure}

To verify the feasibility of substituting expensive \textit{ab initio} calculations with more efficient machine learning methods in phase diagram calculations, we compare free energy calculations across various systems. We start with the simple \ch{Au-Pt} binary system to discuss the influence of formation energy error from SQS introduced by the MLIP. The \ch{Au-Pt} binary phase diagram features a single FCC phase region, with a miscibility gap at low temperature. To capture the spinodal decomposition, particularly the critical temperature, we calculate the formation energies of FCC SQS's at different compositions. The critical temperature of \ch{Au-Pt} miscibility gap is \SI{1533}{\kelvin} according to experimental data \cite{okamoto1985pt}, while SQS and VASP calculations yield a critical temperature of \SI{1661}{\kelvin}. This discrepancy between experiments and DFT-derived thermodynamic properties is common and can be explained by inaccuracies in the DFT calculations as well as potentially ignored contributions to the free energies of phases at elevated temperatures.

\begin{figure}[hbt]
    \includegraphics[width=\columnwidth]{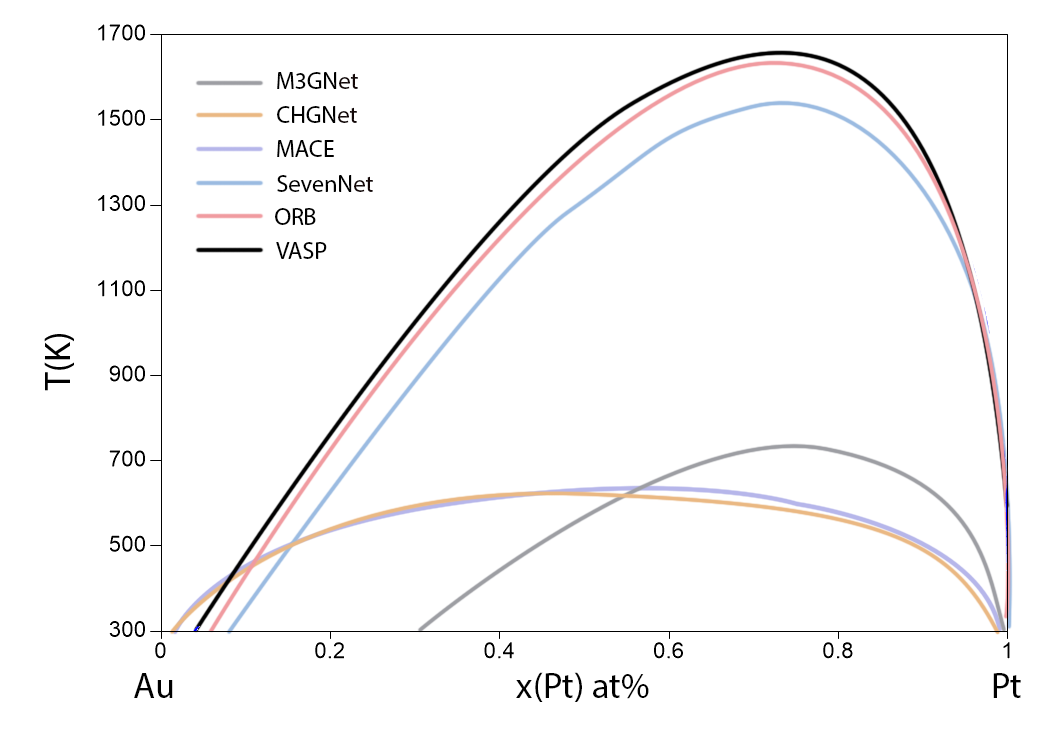}
    \caption{Au-Pt binary phase diagram plotted with VASP and MLIP data.}
    \label{fig:AuPtPD}
\end{figure}

In \Cref{fig:AuPtenergy}, we present the formation energy of \ch{Au-Pt} at \SI{0}{\kelvin}, calculated using VASP, M3GNet, CHGNet, MACE, SevenNet and ORB. Although the absolute error in the total energy appears low (less than \SI{40}{\milli\electronvolt} per atom), the relative errors in formation energy can be significant compared to VASP results, leading to a noticeable underestimation of the critical temperature for spinodal decomposition, as shown in \Cref{fig:AuPtPD}. For M3GNet, CHGNet and MACE, the critical temperature is over \SI{900}{\kelvin}, much lower than the results obtained using VASP. For SevenNet, although the critical temperature is very close to experimental data, it is still \SI{126}{\kelvin} lower than VASP. In this case, the errors in the free energy calculations and those introduced by SQS and CALPHAD coincidentally cancel out. In contrast, ORB's result matches the VASP result closely, with the critical temperature being \SI{32}{\kelvin} higher than VASP, as the average error in formation energies at \SI{0}{\kelvin} is only \SI{1.9}{\percent}. These results indicate that for MLIPs to be useful as proxies for much more expensive DFT calculations, errors in total energies with respect to the DFT ground truth should be in the low \si{\milli\electronvolt} range.

\begin{table*}[t]
    \centering
    \caption{Average Error in formation energy calculations of SQS's and critical temperature of miscibility gap calculated with VASP and MLIPs. "MS" stands for meta-stable and "SP" stands for single phase without miscibility gap.}
    \label{tab:binary_systems}
    \resizebox{\textwidth}{!}{%
    \begin{tabular}{@{}lcccccccccccc@{}}
    \toprule
     & Experiment & VASP & \multicolumn{2}{c}{M3GNet} & \multicolumn{2}{c}{CHGNet} & \multicolumn{2}{c}{MACE} & \multicolumn{2}{c}{SevenNet} & \multicolumn{2}{c}{ORB} \\ \cmidrule(l){2-13} 
     & $T_c$ (\si{\kelvin}) & $T_c$ (\si{\kelvin}) & Error & $T_c$ (\si{\kelvin}) & Error & $T_c$ (\si{\kelvin}) & Error & $T_c$ (\si{\kelvin}) & Error & $T_c$ (\si{\kelvin}) & Error & $T_c$ (\si{\kelvin}) \\ \midrule
    \ch{Cr-Mo} (BCC) & 1153 \cite{venkatraman1987cr} & 2038 & \SI{106.7}{\percent} & 3880 & \SI{107.2}{\percent} & SP & \SI{109.3}{\percent} & SP & \SI{65.6}{\percent} & 1398 & \SI{4.3}{\percent} & 2019 \\
    \ch{Cr-Nb} (BCC) & MS \cite{du2005thermodynamic} & 3590 & \SI{27.1}{\percent} & 3753 & \SI{51.5}{\percent} & 1509 & \SI{11.0}{\percent} & 3148 & \SI{13.4}{\percent} & 2998 & \SI{1.6}{\percent} & 3506 \\
    \ch{Nb-V} (BCC) & 1077 \cite{yang2021experimental} & 1359 & \SI{17.9}{\percent} & 1033 & \SI{26.4}{\percent} & 854 & \SI{22.0}{\percent} & 1598 & \SI{26.5}{\percent} & 1631 & \SI{14.5}{\percent} & 1403 \\
    \ch{Au-Pt} (FCC) & 1533 \cite{okamoto1985pt} & 1661 & \SI{73.8}{\percent} & 740 & \SI{40.9}{\percent} & 618 & \SI{39.3}{\percent} & 642 & \SI{12.3}{\percent} & 1535 & \SI{1.9}{\percent} & 1693 \\
    \ch{Ag-Pt} (FCC) & MS \cite{karakaya1987ag} & 1227 & \SI{161.8}{\percent} & SP & \SI{98.0}{\percent} & 872 & \SI{62.2}{\percent} & 1301 & \SI{84.9}{\percent} & 1543 & \SI{14.2}{\percent} & 1158 \\ \bottomrule
    \end{tabular}%
    }
    \end{table*}

In \Cref{tab:binary_systems}, we present additional results of binary systems as illustrative examples. To ensure a clear comparison between methods, our focus is restricted to the simple BCC/FCC phases exhibiting a miscibility gap, excluding the liquid phase and any potential intermetallic compound phases (e.g., the Laves phase C14 in the \ch{Cr-Nb} system). The error in formation energies is defined as follows:

\begin{equation}
    \text{Error} = \frac{\sum\limits_x{\lvert \Delta G_\text{MLIP}\left(x\right)} - \Delta G_\text{FP}\left(x\right)\rvert}{\sum\limits_x \lvert \Delta G_\text{FP}\left(x\right)\rvert}
\end{equation}

where $\Delta G_\text{MLIP}\left(x\right)$ and $\Delta G_\text{FP}\left(x\right)$ represent the formation free energy at composition $x$ at \SI{0}{\kelvin}, calculated using machine-learning interatomic potentials and DFT, respectively. Machine-learning models like M3GNet, CHGNet, and MACE often have large relative errors in formation energy calculations. These errors can lead to incorrect predictions, such as identifying single-phase regions instead of the expected miscibility gaps in the phase diagrams. SevenNet performs better overall, with smaller errors, but it can still fail for certain systems like \ch{Cr-Mo} and \ch{Ag-Pt}, which are particularly challenging. On the other hand, ORB consistently matches VASP calculations with high accuracy. Across all tested systems, ORB achieves average energy errors below \SI{15}{\percent} and differences in critical temperatures of less than \SI{100}{\kelvin}, making it a reliable tool for predicting phase stability.


Building on the discussion of model accuracy, it is important to consider how errors in formation energy can affect phase diagrams, especially for systems with multiple competing phases. Even small inaccuracies in energy predictions can lead to significant changes in the calculated phase stability. For systems with several competing phases, the errors in the calculated formation (i.e. relative) energies among those phases can gradually influence the phase diagram calculation, as the energy difference between competing phases tends to be rather small. For instance, the total energies of BCC, FCC, HCP, and omega \ch{Hf} at \SI{0}{\kelvin} and \SI{0}{\giga\pascal} differ by less than \SI{0.2}{\electronvolt} per atom, with the energy difference between omega and FCC phases being less than \SI{0.03}{\electronvolt} per atom \cite{zhang2019first}. Consequently, even a small error of \SI{0.05}{\electronvolt} per atom in total energy can alter the predicted stable phase, when considering the competition for phase stability as the result of the minimization of the total Gibbs free energy of the system. Since the prediction errors are not trivially correlated, discrepancies with respect to DFT calculations do not necessarily cancel out and cannot be simply corrected by simple re-scaling approaches. Thus, these propagating, irreducible, errors may lead to significant qualitative differences in the phase stability predictions. This in turn can lead to predicted phase diagrams that are qualitatively and topologically different from predictions that use DFT calculations.

\begin{figure}[H]
     \centering
    \includegraphics[width=\linewidth]{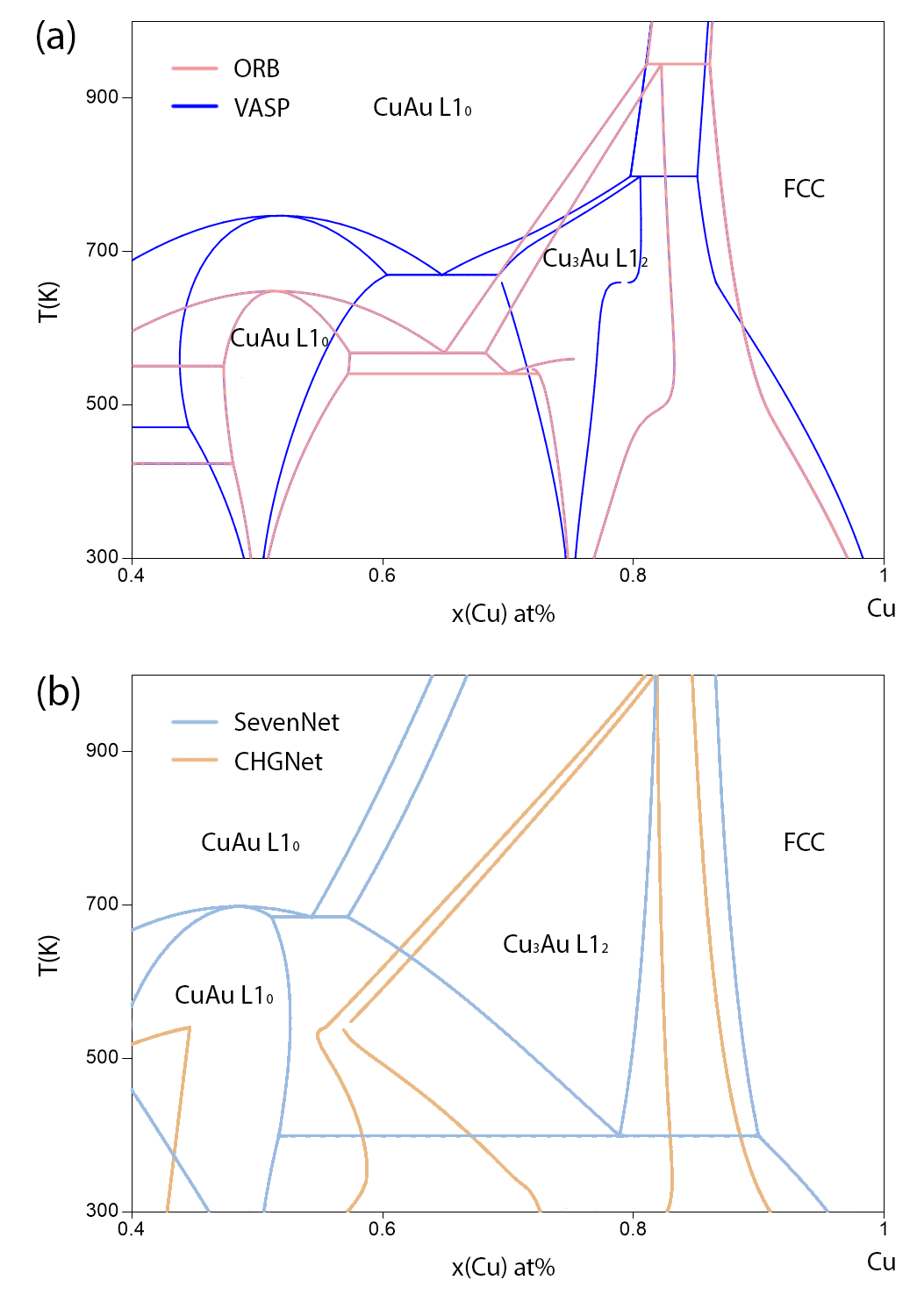}
    \caption{Cu-Au binary phase diagram on the \ch{Cu}-rich side plotted with (a) VASP and ORB; (b) SevenNet and CHGNet}
    \label{fig:CuAu}
\end{figure}

Using the \ch{Au-Cu} binary system as an example, several intermetallic compounds - \ch{CuAu} L1$_0$ phase, \ch{Cu3Au} L1$_2$ phase, and \ch{CuAu3} L1$_2$ phase - compete with the FCC solid solution. As shown in \Cref{tab:CuAu_system}, formation energies of \ch{Cu-Au} intermetallic compounds calculated with CHGNet and ORB are close to VASP results, while the energies calculated with M3GNet, MACE and SevenNet exhibit larger errors. These errors are non-trivial, as they produce phase diagrams with topologies that are radically different from the predicted topologies in the phase diagram using DFT calculations.

In \Cref{fig:CuAu}, we present the \ch{Cu-Au} phase diagrams on the \ch{Cu}-rich side, constructed using calculations from VASP, ORB, SevenNet, and CHGNet. All methods correctly predict the stability of \ch{CuAu} L1$_0$ phase and \ch{Cu3Au} L1$_2$ phase, although they overestimate the stability of L1$_0$ phase at high temperature compared to experimental results \cite{okamoto1987cu}. The phase diagram plotted with ORB, shown in \Cref{fig:CuAu} (a), gives the equilibrium temperature that is \SI{100}{\kelvin} lower than VASP for \ch{CuAu} L1$_0$ phase and \SI{150}{\kelvin} higher for \ch{Cu3Au} L1$_2$ phase. These deviations are within an acceptable range and preserve the overall topology of the phase diagram. \par

In contrast, phase diagrams generated by the other potentials, such as SevenNet and CHGNet in \Cref{fig:CuAu} (b), exhibit larger errors or even fail to predict the correct topology. For example, SevenNet significantly underestimates the formation energy of the \ch{Cu3Au} L1$_2$ phase, calculating it as only half of the VASP result. This error causes the \ch{Cu3Au} L1$_2$ phase to lose its thermodynamic stability at low temperatures. Consequently, the single-phase region of \ch{Cu3Au} L1$_2$ phase is overtaken by the two-phase region of \ch{CuAu} L1$_0$ and FCC below \SI{400}{\kelvin}. At higher temperatures, the \ch{Cu3Au} L1$_2$ phase regains stability due to entropy contributions, but the equilibrium temperature; however, the equilibrium temperature of \ch{Cu3Au} L1$_2$ phase is gradually overestimated. The CHGNet result, on the other hand, underestimates the formation energy of \ch{Cu3Au} L1$_2$ by \SI{282}{\milli\electronvolt} per cell or \SI{71}{\milli\electronvolt} per atom (for ORB, the error is \SI{156}{\milli\electronvolt} per cell or \SI{39}{\milli\electronvolt} per atom), leading to an overestimation of equilibrium temperature by \SI{250}{\kelvin}, and affecting the interaction with \ch{CuAu} L1$_0$ phase, thus distorting the phase boundaries, although it almost perfectly calculates the formation energy of \ch{CuAu} L1$_0$ phase. 

\begin{table*}
    \centering
    \caption{Formation Energies and Errors of Cu-Au system calculated by VASP and MLIPs.}
    \label{tab:CuAu_system}
    \resizebox{\textwidth}{!}{%
    \begin{tabular}{@{}lccccccccccc@{}}
    \toprule
     & VASP & \multicolumn{2}{c}{M3GNet} & \multicolumn{2}{c}{CHGNet} & \multicolumn{2}{c}{MACE} & \multicolumn{2}{c}{SevenNet} & \multicolumn{2}{c}{ORB} \\ \cmidrule(l){2-12} 
     & $\Delta G \text{(eV)}$ & $\Delta G \text{(eV)}$ & Error & $\Delta G \text{(eV)}$ & Error & $\Delta G \text{(eV)}$ & Error & $\Delta G \text{(eV)}$ & Error & $\Delta G \text{(eV)}$ & Error \\ \midrule
    \ch{CuAu} L1$_0$ & -0.1059 & -0.1011 & \SI{-4.6}{\percent} & -0.1035 & \SI{-2.2}{\percent} & -0.1019 & \SI{-3.8}{\percent} & -0.1037 & \SI{-2.0}{\percent} & -0.0869 & \SI{-17.9}{\percent} \\
    \ch{Cu3Au} L1$_2$ & -0.1756 & -0.1327 & \SI{-24.4}{\percent} & -0.1474 & \SI{-16.0}{\percent} & -0.1635 & \SI{-6.9}{\percent} & -0.0772 & \SI{-56.0}{\percent} & -0.1600 & \SI{-8.9}{\percent} \\
    \ch{CuAu3} L1$_2$ & -0.0822 & -0.0517 & \SI{-37.1}{\percent} & -0.0936 & \SI{13.9}{\percent} & -0.0346 & \SI{-58.0}{\percent} & -0.0547 & \SI{-33.5}{\percent} & -0.0609 & \SI{-25.9}{\percent} \\ \bottomrule
    \end{tabular}%
    }
\end{table*}

\begin{figure}[hbt]
    \centering
    \includegraphics[width=\linewidth]{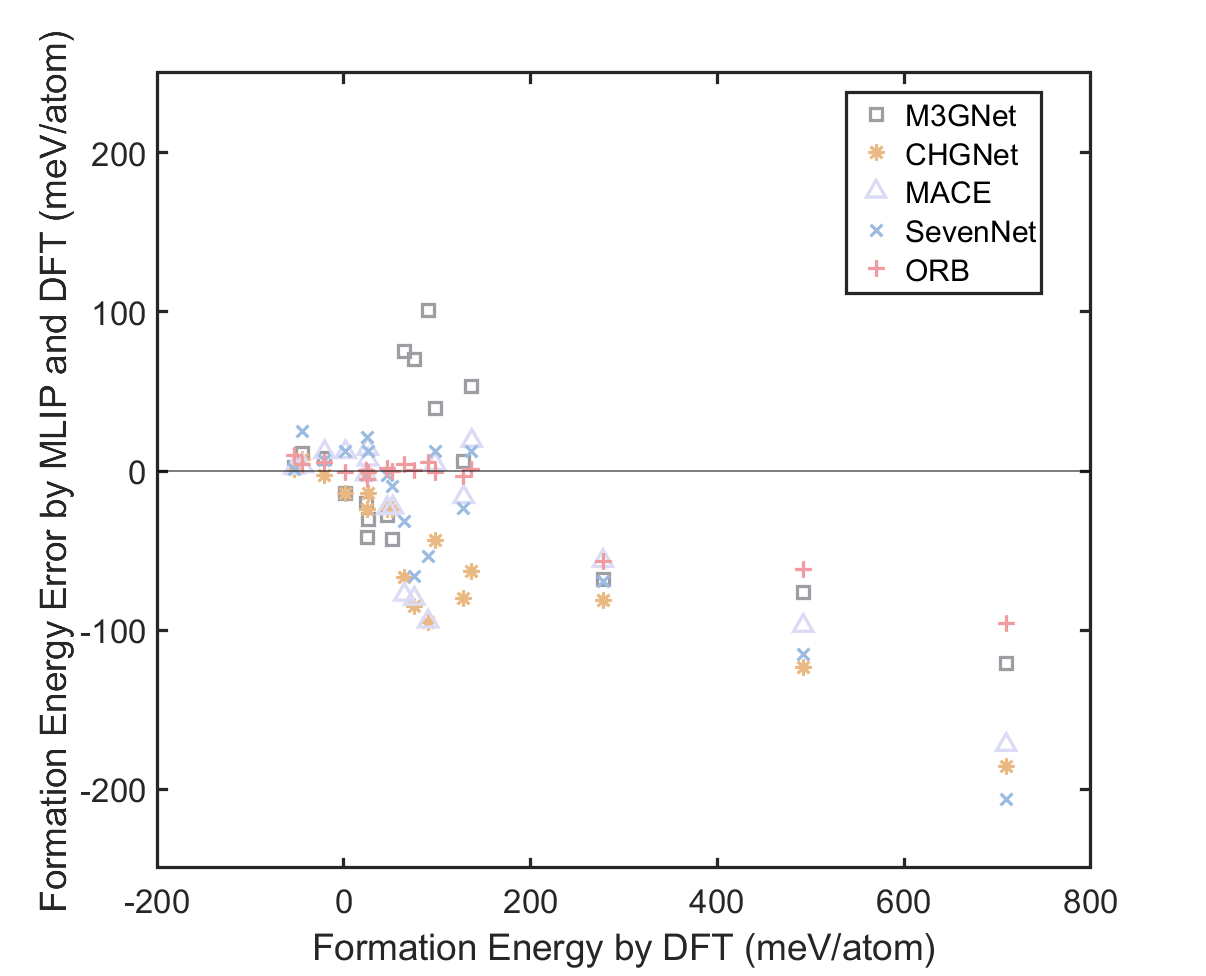}
    \caption{Errors of formation energies calculated with different MLIPs compared to DFT. Data from systems in Table 1 and 2.}
    \label{fig:energycompare}
\end{figure}

In \Cref{fig:energycompare}, we plot all the formation energies we have in \Cref{tab:binary_systems,tab:CuAu_system}, comparing the error of different MLIPs. The figure shows that most of the formation energies calculated with ORB tend to produce results almost in perfect agreement with the VASP calculations, although there are a few structures that exhibit marked discrepancies with the DFT calculations. Even in this cases, the errors tend to be generally smaller than what is obtained using the other potentials. In contrast, the discrepancy between MLIP and DFT-calculated formation energies tends to be generally larger than ORB, with M3GNet producing some of the most significant discrepancies.

Replacing VASP calculations with ORB provides a substantial improvement in computational efficiency. Using the \ch{Cr-Mo} BCC system as an example, we compare the CPU time required for relaxation and free energy calculations, as detailed in \Cref{tab:timecost}. For large SQS structures, ORB can be over 1000 times faster than traditional DFT methods. Although ionic steps for structural relaxation are still necessary, the use of MLIP eliminates the need for the computationally intensive electronic self-consistency steps required in DFT calculations.

\begin{figure}[hbt]
    \centering
    \includegraphics[width=\linewidth]{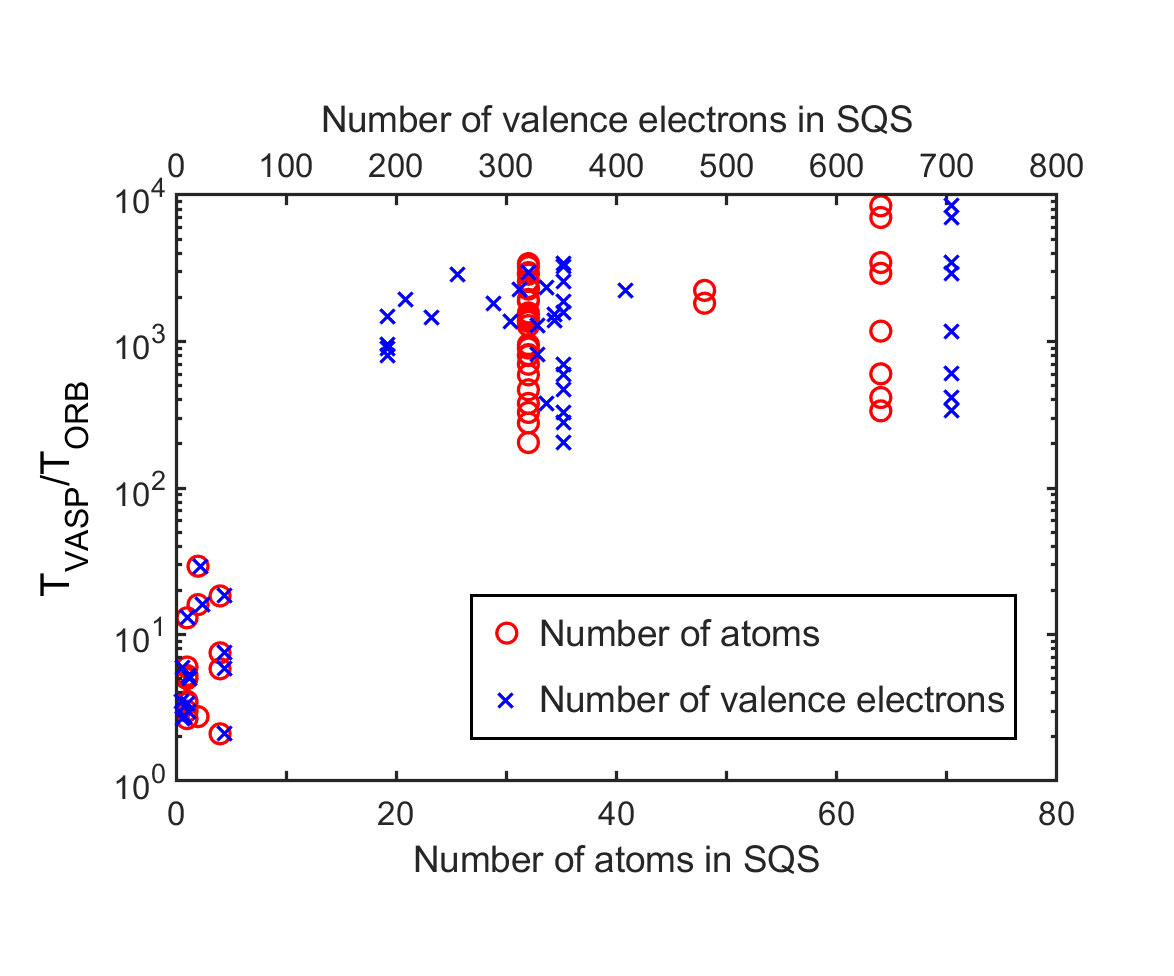}
    \caption{Ratio of time costs between VASP and ORB for SQS Relaxations vs the number of atoms and valence electrons}
    \label{fig:timecost}
\end{figure}

\Cref{fig:timecost} highlights the time cost comparison between VASP and ORB during the relaxation of various SQS structures. The results clearly show that the ratio of VASP time to ORB time increases as the number of atoms or valence electrons in the structure grows. This difference arises because DFT calculations scale approximately as \(N^4\) with the number of atoms \(N\), whereas MLIP-based calculations exhibit much more efficient scaling. As a result, the computational advantage of MLIP-based approaches like ORB becomes increasingly significant for larger systems. This is particularly important for predicting the properties of alloys with complex stoichiometries, where larger supercells are often required to capture the intricate chemical and structural features. By leveraging the efficiency of MLIPs, it is possible to extend the scope of our simulations to more complex systems at much lower costs than when using DFT as the energy calculation engine.

\begin{table*}
    \centering
    \caption{Time cost of \ch{Cr-Mo} BCC system calculated by VASP and MLIPs.}
    \label{tab:timecost}
    \begin{tabular}{c|ccccc}
    \toprule
    & \ch{Cr} & \ch{Cr_{0.75}Mo_{0.25}} & \ch{Cr_{0.5}Mo_{0.5}} & \ch{Cr_{0.25}Mo_{0.75}} & \ch{Mo} \\ \hline
    Number of atoms in SQS & 1 & 32 & 48 & 32 & 1 \\ 
    Total CPU time with VASP (sec) & 8 & 26178 & 69367 & 27972 & 12\\
    Total CPU time with ORB (sec) & 3 & 29 &38 & 19& 2\\
    \bottomrule
    \end{tabular}%
\end{table*}

\begin{table*}
  \centering
  \caption{Vibrational Entropy calculated by ATAT with VASP and ORB (in Boltzmann Constant per cell)}
  \label{tab:entropy}
  \begin{tabular}{@{}lccc@{}}
  \toprule
   &  & VASP & ORB \\ \midrule
  \multirow{2}{*}{FCC A1} & \ch{Cu} & -87.5464 & -87.6627 \\
  \multicolumn{1}{c}{} & \ch{Au} & -85.2416 & -85.5381 \\ \midrule
  \multirow{3}{*}{CuAu L1$_0$} & \ch{CuCu} & -175.030 & -175.327 \\
   & \ch{CuAu} & -172.621 & -172.809 \\
   & \ch{AuAu} & -170.689 & -171.086 \\ \midrule
  \multirow{4}{*}{CuAu GAMMA L1$_2$} & \ch{Cu3Cu} & -350.147 & -350.647 \\
   & \ch{Cu3Au} & -347.833 & -347.722 \\
   & \ch{Au3Cu} & -343.383 & -343.214 \\
   & \ch{Au3Au} & -341.327 & -342.020 \\ \bottomrule
  \end{tabular}%
\end{table*}

At high temperatures, thermally excited degrees of freedom need to be accounted for in order to have a more accurate description of the phase stability of competing phases. Here, we use the ORB potential to calculate the vibrational entropy (and corresponding free energy). We calculate the vibrational entropy of FCC, L1$_0$ and L1$_2$ phase in \ch{Au-Cu} system in the high-temperature limit, as shown in \Cref{tab:entropy}. The close agreement between the results indicates that ORB can reliably calculate vibrational entropy, supporting the accuracy of phase diagram predictions. This capability is important when considering materials systems that operate at high temperatures---such as refractory high entropy alloys, RHEAS---as these contributions tend to be more significant as temperature increases \cite{vela2023high}.\par

A major challenge when attempting to construct reliable thermodynamic descriptions of alloys is the estimation of the thermodynamic properties of the liquid phase. These properties can only be obtained through expensive \textit{ab initio} calculations \cite{hong2016user} with computational cost orders of magnitude higher than the costs associated with the computation of thermodynamic properties of crystals in their ground state. Here, having established its general reliability across many systems, we have used the ORB MLIP to carry out molecular dynamics calculations of alloys in the liquid state. As an example, we consider the \ch{Pt-W} binary system. We first generated liquid SQS configurations using ATAT at various concentrations for the liquid phase. In the molecular dynamics simulations, we employ the NPT ensemble for the first 3,000 steps with a time step of \SI{1}{\femto\second} to obtain an appropriate structure and cell size. This is followed by 10,000 steps in the NVT ensemble to compute the average total energy. The temperature is set \SI{50}{\kelvin} above the melting points of \ch{Pt} and \ch{W}, with linear interpolation applied for the mixed structures. \par

\begin{figure}[hbt]
    \centering
        \includegraphics[width=\linewidth]{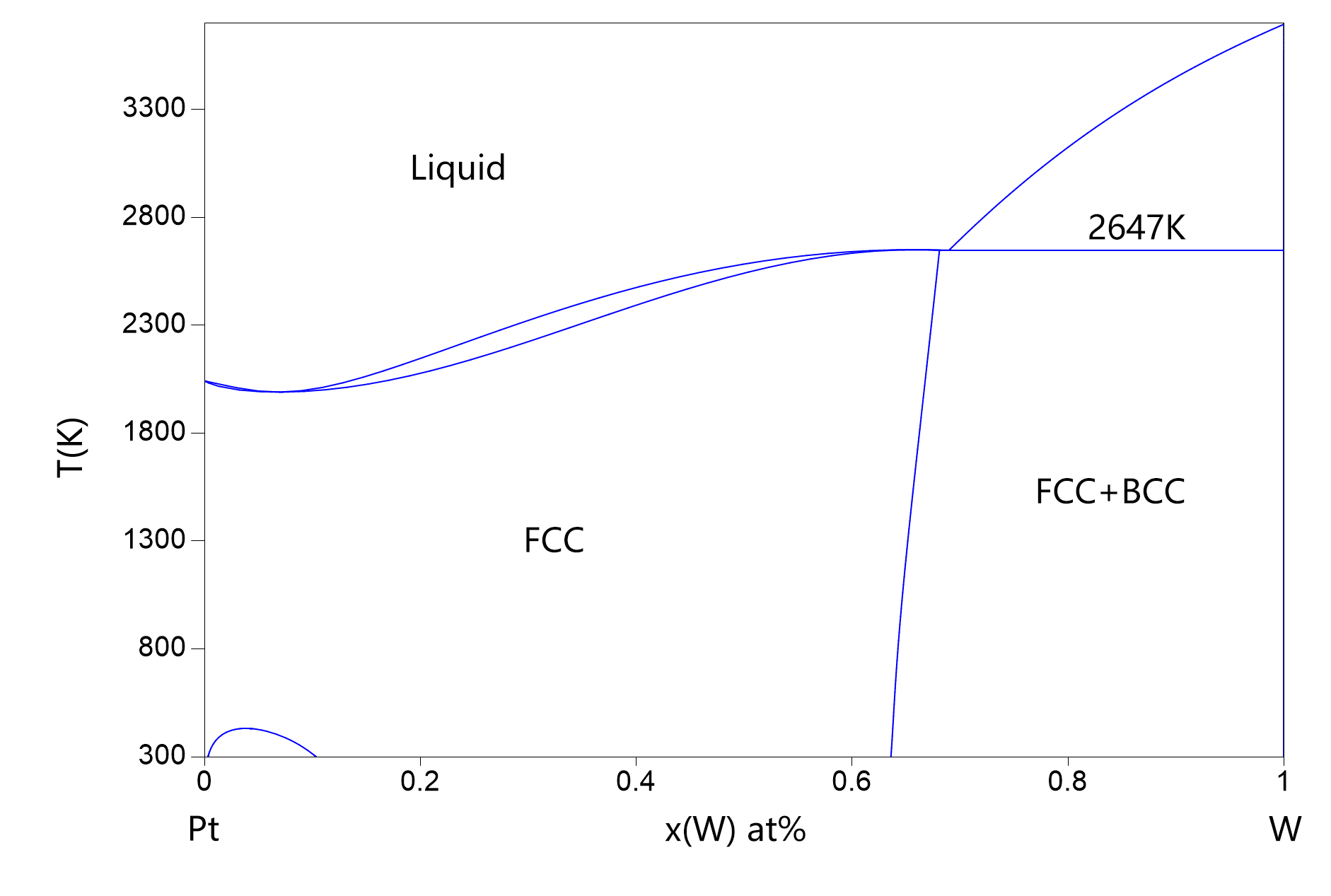}
    \caption{Pt-W binary phase diagram calculated with ORB. MD is used for the liquid phase, and SQSs are used for the solid phase.}
    \label{fig:PtW}
\end{figure}

Using the total energies derived from MD simulations with ORB, along with data for BCC and FCC solid phases, we calculate the phase diagram of the \ch{Pt-W} system, shown in \Cref{fig:PtW}. The phase diagram shows that the \ch{Pt-W} system has a broad FCC phase region on the \ch{Pt}-rich side, while BCC \ch{W} exhibits almost no solubility for \ch{Pt}, in agreement with experiments. The equilibrium temperature of the eutectic phase diagram is \SI{2647}{\kelvin}, which is only \SI{86}{\kelvin} higher than experimental results \cite{hansen1958constitution,knapton1980alloys,guillermet1995phase}. We note that the calculation of this phase diagram using entirely \textit{ab initio} methods would be many orders of magnitude more costly than what was achieved using this efficient and relatively accurate machine learning potential. 

\begin{figure}[hbt]
    \centering
        \includegraphics[width=\linewidth]{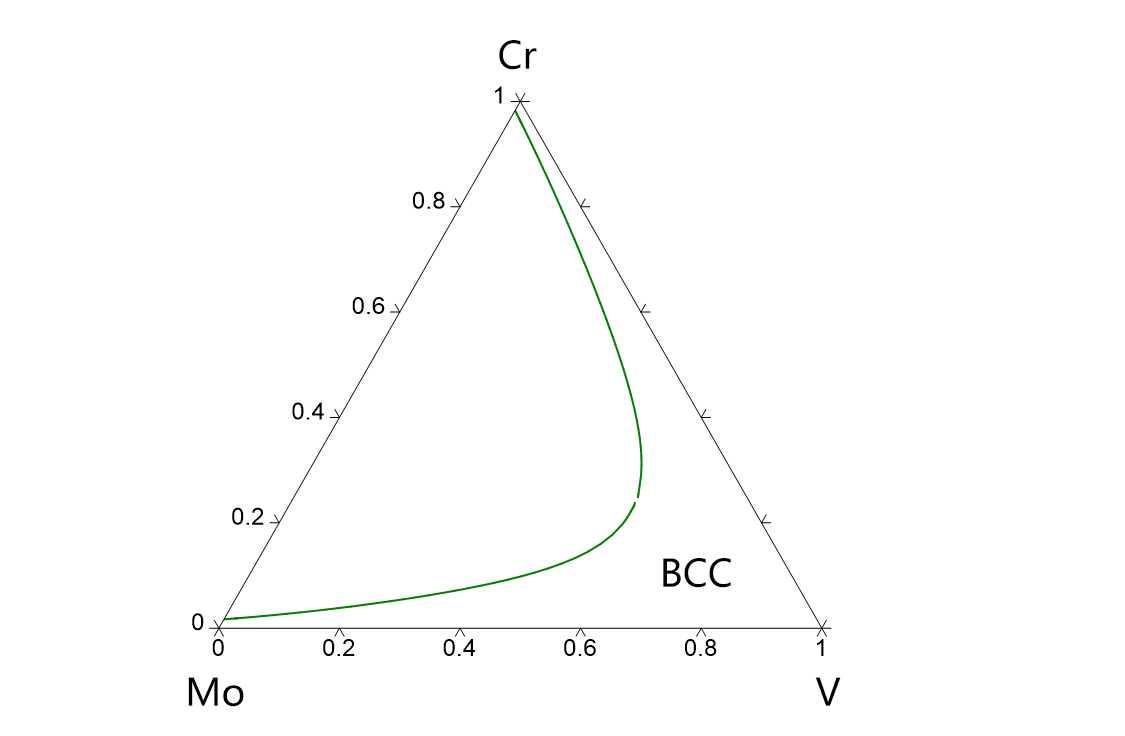}
    \caption{\ch{Cr-Mo-V} ternary phase diagram at \SI{700}{\degreeCelsius} calculated with ATAT and ORB potential.}
    \label{fig:CrMoV}
\end{figure}

The benefits of using MLIPs to obtain reliable and accurate---at the DFT level---CALPHAD descriptions of multi-component, multi-phase systems extend beyond binary systems. In \Cref{fig:CrMoV}, we present a ternary phase diagram for the \ch{Cr-Mo-V} system, calculated using the ORB method. Compared to previous results obtained with the TCHEA4 database and cluster expansion \cite{zhu2023probing}, the calculated phase diagram accurately predicts the miscibility gap of the BCC phase on the \ch{Cr-Mo} side and provides a reliable phase diagram overall. This result demonstrates that our method is well-suited for exploring high-entropy alloys, as the entire computational process for the \ch{Cr-Mo-V} system required less than 2 hours using sequential computation on one core on a supercomputer. Importantly, since the calculations were fitted to a CALPHAD model, it is possible to carry out extrapolations and predictions over broad temperature ranges, provided other potential competing phases---possibly stabilized at lower temperatures---are included in the analysis.

\section{Conclusion}

In this study, we demonstrated the potential of machine learning interatomic potentials (MLIPs) as an efficient alternative to expensive \textit{ab initio} calculations for alloy phase diagram modeling using Special Quasirandom Structures (SQS). While total energy errors for MLIPs are typically below \SI{0.1}{\electronvolt/atom} compared to first-principles results, their impact on the relative formation enthalpy can significantly affect phase diagram accuracy, leading to shifts in phase boundaries and occasional topology errors. Among the MLIPs evaluated, ORB or future potentials with equal or better accuracy emerged as the most reliable, achieving an absolute total energy error below \SI{10}{\milli\electronvolt/atom}, which is critical for maintaining accurate phase stability predictions. Additionally, vibrational entropy calculations and molecular dynamics simulations with ORB align closely with \textit{ab initio} results, further validating its utility for thermodynamic modeling.

Despite minor errors, the remarkable efficiency of MLIPs, offering speedups of several orders of magnitude compared to traditional methods, makes them highly valuable for high-throughput calculations. Phase diagram calculations that traditionally require days or weeks with DFT can now be completed in under an hour with MLIPs, even for complex binary systems. This efficiency enables large-scale exploration of alloy phase spaces, particularly for compositionally complex systems like high-entropy alloys, facilitating rapid discovery of novel phases and materials. Furthermore, phase diagrams calculated using MLIPs can serve as initial guesses or references for active learning frameworks \cite{zhu2024active, deshmukh2024active,dai2020efficient,ament2021autonomous,kusne2020fly}, enabling iterative refinement with experimental data. This hybrid approach could help compensate for the inaccuracies in MLIP-based phase diagrams, ultimately bridging the gap between computational and experimental methods.

Looking forward, several important questions remain unanswered:
\begin{itemize}
    \item Can ORB or future potentials with equal or better accuracy maintain high performance across diverse alloy systems, including those with non-metallic elements, or would further training or modifications be required?
    \item What specific factors contribute to the superior performance of ORB or similar potentials compared to other MLIPs—model architecture, training dataset, or both? How might these insights guide the development of even more accurate MLIPs?
    \item Given the computational efficiency of MLIPs, how can the SQS method be refined or replaced to reduce approximations, enabling even more accurate representation of disordered atomic structures?
\end{itemize}

An additional critical avenue for future research involves extending high-throughput computational methods to develop CALPHAD descriptions for high-entropy materials beyond high-entropy alloys. The high-entropy paradigm, originally focused on metallic systems, has now expanded into a broader materials space, including ceramics, semiconductors, and other multi-component single-phase materials \cite{gao2018high,guo2024role,schweidler2024high,tang2024rapid}. These materials are being actively explored for applications in areas such as battery technologies, catalysis, and magnetic systems, among others. Despite the growing interest and rapid advancements in these fields, there are currently no reliable CALPHAD databases for these systems. Without such CALPHAD descriptions, the exploration of these vast materials spaces would be akin to trying to find, blindfolded, a needle in a (multi-dimensional) haystack. Experimental approaches to phase stability in such multi-component systems are prohibitively expensive, underscoring the necessity for approximate but scalable methods based on DFT calculations.

Developing high-throughput workflows that combine the efficiency of MLIPs with the accuracy of DFT-derived thermodynamic data could provide a practical pathway to address this challenge. These approaches would enable the systematic construction of CALPHAD databases for a wide range of high-entropy materials, facilitating the exploration of complex multi-phase systems and accelerating the discovery of new functional materials for advanced applications. By addressing these challenges and embracing these promising future research directions, MLIP-based methods and high-throughput CALPHAD workflows can significantly advance materials design, paving the way for innovative solutions in alloy design and beyond.
\section*{Acknowledgements}
The authors would like to acknowledge the support of the National Science Foundation through Grant No. 2119103. We also acknowledge the support from the Army Research Office through Grant No. W911NF-22-2-0117. Calculations were carried out at Texas A\&M High-Performance Research Computing (HPRC) Facility.

\bibliographystyle{unsrt}
\bibliography{bib}
\end{document}